\documentclass[aps,pra,superscriptaddress,amsmath,amsfonts,amssymb,floatfix,nofootinbib]{revtex4}
\usepackage{enumerate}
\usepackage{mathtools}
\usepackage{amsmath}
\usepackage{graphicx}
\usepackage{epsfig}
\usepackage{sidecap}
\usepackage{hyperref,stackrel}
\usepackage[normalem]{ulem}
\usepackage{color}
\usepackage{enumerate}
\usepackage{bm}
\usepackage[utf8]{inputenc}
\usepackage{amsmath}
\usepackage{amsfonts}
\usepackage{amssymb}
\usepackage{graphicx}
\usepackage{enumitem}
\usepackage{physics}

\begin{document}

\title{Open quantum system dynamics of $X$-states: Entanglement sudden death and sudden birth}
\author{Nikhitha Nunavath} \thanks{nikhithanunavath@students.iisertirupati.ac.in}
\affiliation{Jaypee Institute of Information Technology, A-10, Sector-62, Noida, UP-201309, India}

\author{Sandeep Mishra} \thanks{sandeep.mtec@gmail.com }
\affiliation{Jaypee Institute of Information Technology, A-10, Sector-62, Noida, UP-201309, India}

\author{Anirban Pathak} \thanks{anirban.pathak@gmail.com}
\affiliation{Jaypee Institute of Information Technology, A-10, Sector-62, Noida, UP-201309, India}

\begin{abstract}
The origin of disentanglement for two specific sub-classes of $X$-states namely  maximally nonlocal mixed states (MNMSs) and maximally entangled mixed states (MEMSs) is investigated analytically for a physical system consisting of two spatially separated qubits interacting with a common vacuum bath. The phenomena of entanglement sudden death (ESD) and the entanglement sudden birth (ESB) are observed, but the characteristics of ESD and ESB are found to be different for the case of two photon coherence and single photon coherence states. The role played by initial coherence for the underlying entanglement dynamics is investigated.  Further, the entanglement dynamics of MNMSs and MEMSs under different environmental noises namely phase damping, amplitude damping and RTN noise with respect to the decay and revival of entanglement is analyzed. It's observed that the single photon coherence states are more robust against the sudden death of entanglement indicating the usability of such states in the development of technologies for the practical implementation of quantum information processing tasks.  
\end{abstract}
\maketitle

\section{Introduction}
Quantum entanglement is an important nonclassical phenomenon that plays a key role in the realization of various tasks related to quantum computing, quantum communication and quantum metrology \cite{nielsen2000quantum}. In fact, entanglement lies at the heart of many phenomenal applications ranging from quantum teleportation \cite{bennett1993teleporting} to quantum computation and quantum key distribution (QKD) \cite{horodecki2009quantum,shenoy2017quantum,xu2020secure,pirandola2020advances} to atomic and molecular spectroscopy \cite{bollinger1996optimal,huelga1997improvement}. It is important for a quantum system to retain its properties under the evolution for its experimental implementations. For the practical realizations, it's extremely important to master the properties of entanglement at the level of single atoms and photons to store and transfer information with high fidelity. In a case like bipartite quantum system, when the quantum state evolves under certain environmental dynamics, the system undergoes interaction with the environment leading to the decoherence and degradation of entanglement in a finite amount of time \cite{aolita2015open}. This decoherence in quantum system can lead to sudden loss of entanglement and thus causing the entanglement sudden death (ESD)  \cite{yu2004finite, yu2006quantum, yu2009sudden}. The entanglement sudden death (ESD) was experimentally confirmed with photonic qubits and atomic ensemble \cite{salles2008experimental,laurat2007heralded,almeida2007environment}. The ESD is undesirable for the entangled states used in the implementation of QKD, quantum teleportation and quantum computation protocols as they require a steady entanglement over time. Further, it has also been shown that in many situations, such as  spontaneous emission from two spatially separated atoms, interacting with bath like vacuum reservoir \cite{ficek2002entangled,tanas2003entanglement, tanas2004entangling,jakobczyk2002entangling}, atoms driven by laser field \cite{tanas2003entanglement}, and atoms coupled to squeezed vacuum \cite{tanas2004stationary} leads to ESD. 

Moreover, the phenomenon of revival of entanglement in finite time, followed by the ESD has also been observed for many initially unentangled qubits \cite{ficek2002entangled,ficek2010spontaneously,tanas2003entanglement,ficek2010sudden}. The spontaneous evolution of separable qubits dynamically leads to creation of entanglement  and is termed as entanglement sudden birth (ESB) \cite{ficek2008delayed}. The birth and death of entanglement is highly dependent on the threshold coherence during the evolutionary dynamics of the qubits \cite{ficek2006dark}.  During the time evolution of an entangled state, ESD has been observed in different scenarios (for a review see \cite{orszag2010coherence}). In fact, in open quantum systems, the ESD and ESB are commonly observed features in the studies related to entanglement dynamics. The study of entanglement dynamics for a wide variety of the mixed states and its applications in various fields has now become a very active and important area of research for its practical implementations \cite{cui2009non, tahira2011gaussian, farias2012experimental, mishra2019comparing, mishra2022attainable,wang2018observation,mishra2018probing,thapliyal2017quantum,namitha2018role,li2020entanglement,martin2020entanglement,mortezapour2018coherence,wybo2020entanglement,khedif2018local,amico2018dissipative}. Here, we aim to complement the existing results by investigating open quantum system dynamics of $X$-states with a focus on two sub-classes of $X$-states, namely, maximally nonlocal mixed states (MNMSs) and maximally entangled mixed states (MEMSs). Before we proceed further, it will be apt to state the importance of $X$-states in the domain of quantum information processing to provide the rationale behind selecting these states for the present study. 

Yu and Eberly \cite{yu2004finite} introduced  $X$-state as a specific type of quantum states of interest. Specifically, the $X$-state is viewed as a quantum state having density matrix with non-zero terms only along diagonal and anti-diagonal of the matrix. Thus, in a $X$-state, non-zero terms of the density matrix creating a resemblance with the letter "X" justifying the name. These states have been analysed for the possibility of finite-time disentanglement of two-qubits due to spontaneous emission resulting in ESD \cite{yu2009sudden}. It is important to mention that $X$-states can retain their form even after the time evolution for most of the well-known cases in open quantum systems \cite{quesada2012quantum}. Since their introduction, $X$-states have been recognized as an important class of pure and mixed states, as $X$-states include several sub-classes of quantum states of practical relevance. Specifically, $X$-states  contain maximally entangled states, partially entangled states like Werner states \cite{werner1989quantum}, maximally nonlocal mixed states (MNMSs) \cite{batle2011nonlocality}, maximally entangled mixed states (MEMSs) \cite{munro2001maximizing, ishizaka2000maximally, verstraete2001maximally} and the non-entangled states. Further, the property of $X$-states, allows one to map every two-qubit state to an equivalent $X$-state having same spectrum of entanglement {\cite{mendoncca2014entanglement}}. All these features along with the fact that the above mentioned $X$-states are frequently used as important resources for the realization of various tasks related to quantum communication and computation \cite{altepeter2003ancilla,ali2010quantum,hu2013relations,slaoui2018dynamics,haddadi2020exploring,balthazar2021spin}, have motivated us to conduct the present study.

Continuing with the discussion on the relevance of $X$-states, we may note that among all the mixed states, Werner states \cite{werner1989quantum}, form a convex sum of maximally entangled pure states and maximally mixed states. In general, a Werner state for the bipartite system can be written as 
\begin{equation}
    \rho_{W} = (\frac{1-p}{4}) I_{4} + p\ket{M}\bra{M}
\end{equation}
with $0\leq p\leq1$ and $\ket{M}$ be one of the four  Bell-states, which are maximally entangled. When Werner states interact with the vacuum bath, the phenomena of ESD and ESB are observed. For this, the dynamics of entanglement was investigated for the different classes of coherence \cite{namitha2018role}. Other than Werner states,  MNMS and MEMS also form an important class with applications in many areas of quantum information. These states have their own advantages due to their special structure and evolutionary behavior. MNMSs are useful in teleporting a qubit with fidelity greater than the classical limit under noisy channels \cite{horodecki1996teleportation}. Further, MNMSs remain advantageous for device independent quantum key distribution as they violate Bell inequality \cite{vazirani2019fully,lo2012measurement}. For the case of MEMSs, these states posses the highest possible entanglement for a given amount of mixedness \cite{munro2001maximizing} and also used for ordering problems for mixedness \cite{wei2003maximal}. In short, the MNMS and MEMS are two well-known families of mixed states and have important applications in quantum information processing. Now, as mentioned before, quantum states are fragile and they undergo decoherence under the effect of environment \cite{breuer2002theory}. Furter, the unique features of quantum states such as entanglement undergo degradation and may even lead to ESD \cite{yu2009sudden,almeida2007environment,salles2008experimental}. The effect of the environment on the MNMS and MEMS and thus leading them to ESD will severely affect their application  in the quantum information theoretic tasks. Hence, it's desired to undertake a  carefully study of the entanglement properties of these states under the presence of realistic environments so that one can have complete experimental control over the states during the practical applications. So in this work, we investigated the dynamics of MEMS and MNMS for the one-photon coherence and two-photon coherence of various initial states evolving for the case of two level atom model under the presence of vacuum bath. We further employed the noise to the two-qubit system along with the vacuum bath to study the possibilities of the occurrence of ESD and ESB. We illustrate the dynamical behavior of these states by studying entanglement via the use of concurrence. This study will provide valuable insights for exploring techniques towards the development of quantum systems which exploit quantum entanglement for the information processing tasks.  

The rest of this paper is organised as follows. In Section \ref{model} we describe the theoretical model considered for the two qubit system. Then, we define the measure of entanglement for $X$-states in section \ref{concurrence}. Next, in Section \ref{dynamics}, we analyze the evolutionary dynamics for the MNMS and MEMS for the system under consideration. Then in Section \ref{noise}, we describe the two level atom model under the presence of phase dissipative, amplitude damping and random telegraph channels. Finally, we conclude by summarising the results in Section \ref{conclusion}.

\section{Two atom Model} \label{model}

Since, we are interested in studying the entanglement dynamics of $X$-states with a special emphasis on MNMS and MEMS, so we consider a system of two spatially separated identical two-level atoms, where each atom can be considered as a qubit. Further, we consider the interaction of this system with an environment consisting of a vaccum bath. Here, we aim to study the decoherence dynamics of these two qubits in the Markovian approximation \cite{breuer2002theory}. These distinguishable atoms modelled as qubits when acted upon by transition dipole moment interacting with each other and are coupled to a common vacuum field. The ground state and the excited state of this model can be denoted by $\ket{g_i}$, $\ket{e_i}$ for $i = 1,2$. These states form the  computational basis for the atomic model. The interaction of two qubit system with vacuum bath results in spontaneous emission and the dynamics for this can be modelled using the Lehemberg-Agarwal master equation under Markovian approximation \cite{lehmberg1970radiation,ficek1987quantum}. The time evolution of this two atom model interacting with the environment can be given by the following master equation
\begin{equation}\label{Eqn.2}
    \frac{d\rho}{dt} = -i\sum_{i=1}^{2}\omega_i[S_{i}^{z},\rho]-i\Omega_{ij}\sum_{i\neq j}^{2}[S_{i}^{+}S_{j}^{-},\rho]-\frac{1}{2}\sum_{i,j =1}^{2}\Gamma_{ij}(\rho S_{i}^{+}S_{j}^{-}+S_{i}^{+}S_{j}^{-}\rho - 2S_{j}^{-}\rho S_{i}^{+}),
\end{equation}
where $i\in{1,2}$, $S_{i}^{z} = \ket{e_i}\bra{e_i} - \ket{g_i}\bra{g_i}$ is the
energy operator of the $i$th atomic qubit while $S_{j}^{+} = \ket{e_i}\bra{g_i}$ and $S_{j}^{-} = \ket{g_i}\bra{e_i}$ are the dipole raising and lowering operators respectively. The transition frequencies of the two qubits are considered to be identical, and is taken as $\omega_{i=1,2} = \omega_{0}$. The spontaneous emission rate $\Gamma_{ij}$ for $(i = j)$ is equal
to $\Gamma$. $\Gamma_{ij}$ and $\Omega_{ij}$ ($i\neq j$) represent the
emission rates and the interatomic coupling parameters of two level system, respectively and can be expressed as
\begin{eqnarray}
  \Gamma_{ij} &=& \Gamma_{ji} = \frac{3}{2}\Gamma \Bigg{\{}\Big[1-(\hat{\mu}.\hat{r}_{ij})^2\Big]\frac{\sin({kr_{ij}})}{kr_{ij}} + \Big[1-3(\hat{\mu}.\hat{r}_{ij})^2\Big]\Big{[}\frac{\cos({kr_{ij}})}{(kr_{ij})^{2}}-\frac{\sin(kr_{ij})}{(kr_{ij})^{3}}\Big{]}\Bigg{\}}, \\
\Omega_{ij} &=& \frac{3}{4}\Gamma \Bigg{\{}-\Big[1-(\hat{\mu}.\hat{r}_{ij})^2\Big]\frac{\cos({kr_{ij}})}{kr_{ij}} + \Big[1-3(\hat{\mu}.\hat{r}_{ij})^2\Big]\Big{[}\frac{\sin({kr_{ij}})}{(kr_{ij})^{2}}+\frac{\cos(kr_{ij})}{(kr_{ij})^{3}}\Big{]} \Bigg{\}},
\end{eqnarray}
respectively, where $\hat{\mu}$ is the unit vector along the direction of dipole moment of the two qubit system, $\hat{r}_{ij}$ is the
unit vector along $\Vec{r}_{ij} = |\Vec{r}_{i} - \Vec{r}_{j}|$ and $k = \frac{\omega_{0}}{c}$.\\
The isolated non-interacting two qubit system can be represented in the computational basis given by $\{\ket{g_1}\ket{g_2}, \ket{g_1}\ket{e_2}, \ket{e_1}\ket{g_2}, \ket{e_1}\ket{e_2}\}$. Whereas, the interaction with bath and the dipole-dipole interactions between the identical qubits make the system behave collectively and forms a single four level atomic model. The basis for this system is given by Dicke basis \cite{dicke1954coherence} and is represented for the two qubit system as
\begin{eqnarray} \label{dicke_basis}
\ket{g} &=& \ket{g_1}\ket{g_2}, \\ \nonumber
 \ket{s} &=& \frac{1}{\sqrt{2}}\Big\{\ket{e_1}\ket{g_2} + \ket{g_1}\ket{e_2}\Big\},  \\ \nonumber
  \ket{a} &=& \frac{1}{\sqrt{2}}\Big\{\ket{e_1}\ket{g_2} - \ket{g_1}\ket{e_2}\Big\},  \\ \nonumber
   \ket{e} &=& \ket{e_1}\ket{e_2}.
\end{eqnarray} 
These states form the basis for the collective two qubit model where $\ket{g}, \ket{e}$ are ground and excited states, respectively, while $\ket{s}, \ket{a}$ being the symmetric and anti-symmetric states formed at intermediate level which are maximally entangled.

For this model of two qubit system, we study the entanglement dynamics in the collective basis for the mixed and pure states in the following sections.

\section{Concurrence - Measure of Entanglement} \label{concurrence}
In this section, we use concurrence as  the measure of entanglement for the system under consideration. For, bipartite qubits, concurrence introduced by Wootter \cite{wootters1998entanglement}, which serves as a good measure of entanglement, can be conveniently used for the study of entanglement dynamics. The analytical expression for the concurrence of a two qubit mixed state is given by
\begin{equation}
    C = \rm{\bm{max}} \, \{0,\sqrt{\lambda_{1}}- \sqrt{\lambda_{2}}-\sqrt{\lambda_{3}}-\sqrt{\lambda_{4}}\},
\end{equation}
where $\lambda^{'s}$ are the eigenvalues of $\rho\Tilde{\rho}$ in the descending order and it takes only positive values. The density
matrix $\Tilde{\rho}$ can be written as
\begin{equation}
    \Tilde{\rho} = \sigma_{y}^{A}\otimes\sigma_{y}^{B}\rho^{*}\sigma_{y}^{A}\otimes\sigma_{y}^{B},
\end{equation}
where $\rho^{*}$ is the complex conjugate of the density matrix $\rho$ and $\sigma_{y}$ being the standard Pauli matrix. The values of $C$ are zero for separable states and $0 < C \leq 1$ for entangled states.  

Theoretically, “$X$-states” are represented as the density matrix {in computational basis} containing only non-zero elements in an “$X$” form, along the main diagonal and anti-diagonal sides. The initial form of the $X$-state density matrix can be represented in the computational basis as {($\ket{1}=\ket{g_1}\ket{g_2},\ket{2}=\ket{g_1}\ket{e_2},\ket{3}=\ket{e_1}\ket{g_2},\ket{4}=\ket{e_1}\ket{e_2}$)} as
\begin{equation}
\rho^{AB} =
\begin{bmatrix}
\rho_{11} & 0 & 0 & \rho_{14}\\
0 & \rho_{22} & \rho_{23} & 0\\
0 & \rho_{32} & \rho_{33} & 0\\
\rho_{41} & 0 & 0 & \rho_{44} 
\end{bmatrix},
\end{equation}
with $\rho_{11}+\rho_{22}+\rho_{33}+\rho_{44} = 1$. These $X-$states appear naturally in various scenarios  \cite{pratt2001qubit, bose2001subsystem}. We note that the MNMS and MEMS form a special class of $X$-states. Further, the $X$-states have property that they retain the $X$ form under the noisy evolution \cite{quesada2012quantum}. This robust behavior of $X-$states can be verified by computing concurrence at different time intervals of evolution. For the time evolution of the system, we deal with the collective basis (Eq. \ref{dicke_basis}) which is obtained by unitary transformation of computational basis and is written as 
\begin{equation}
\rho^{AB}(t) =
\begin{bmatrix}
\rho_{gg}(t) & 0 & 0 & \rho_{ge}(t)\\
0 & \rho_{ss}(t) & \rho_{sa}(t) & 0\\
0 & \rho_{as}(t) & \rho_{aa}(t) & 0\\
\rho_{eg}(t) & 0 & 0 & \rho_{ee}(t) 
\end{bmatrix},
\end{equation}
where the explicit forms of the elements \cite{ficek2010sudden} of the above matrix are as follows
\begin{equation}
\begin{split}
    \rho_{ee}(t) & = \rho_{ee}(0)e^{-2\gamma t}, \\
    \rho_{ss}(t) & = \rho_{ss}(0)e^{-(\gamma+\gamma_{12}) t} + \rho_{ee}(0)\frac{\gamma+\gamma_{12}}{\gamma-\gamma_{12}}\Big[e^{(\gamma-\gamma_{12})t }- 1\Big]e^{-2\gamma t},\\
    \rho_{aa}(t) & = \rho_{aa}(0)e^{-(\gamma-\gamma_{12}) t} + \rho_{ee}(0)\frac{\gamma-\gamma_{12}}{\gamma+\gamma_{12}}\Big[e^{(\gamma+\gamma_{12})t }- 1\Big]e^{-2\gamma t},\\
    \rho_{gg}(t) & = 1 - \rho_{ee}(t) - \rho_{ss}(t) - \rho_{aa}(t), \\
    \rho_{ge}(t) & = \rho_{ge}(0)e^{-\gamma t}, \\
    \rho_{sa}(t) & = \rho_{sa}(0)e^{-(\gamma + 2i\Omega_{12})t}.
\end{split}
\end{equation}
Clearly, the diagonal elements of the above density matrix constitute the population terms and the anti-diagonal elements constitute the coherence of the system in the collective basis. Now, the concurrence of a density matrix $\rho^{AB}(t)$ written in standard computational basis can be computed as
\begin{equation}\label{con_fun}
  C(t) = \rm{\bm{max}} \, \{0,C_1(t),C_2(t) \},   
\end{equation}
with,
\begin{equation}
\begin{split}
C_1(t) & =      2\Big{\{}|\rho_{14}(t)|-\sqrt{\rho_{22}(t)\rho_{33}(t)}\Big{\}} \\
    & = 2 \Bigg{\{}|\rho_{ge}(t)| - \sqrt{\Big{[}\frac{\rho_{ss}(t) + \rho_{aa}(t)}{2}\Big{]}^2 - [\rm{Re} \rho_{sa}(t)]^2  }          \Bigg{\}},
\end{split}    
\end{equation}
and
\begin{equation}
\begin{split}
C_2(t) & =      2\Big{\{}|\rho_{23}(t)|-\sqrt{\rho_{11}(t)\rho_{44}(t)}\Big{\}} \\
    & = 2 \Bigg{\{} \sqrt{\Big{[}\frac{\rho_{ss}(t) - \rho_{aa}(t)}{2}\Big{]}^2 + [\rm{Im}  \rho_{sa}(t)]^2 } - \sqrt{\rho_{ee}(t)\rho_{gg}(t)} \Bigg{\}}.
\end{split}    
\end{equation}
These explicit expressions of concurrence provides us the essential mathematical structure required for the rest of the study, where one photon coherence and two photon coherence will be studied systematically for the MNMS and MEMS.

\section{Entanglement dynamics of one photon coherence and two photon coherence of MNMS and MEMS states} \label{dynamics}

In this section, we will study the entanglement dynamics of  MNMS and MEMS for the case of one photon coherence and two photon coherence. It is important to mention the difference between one photon coherence and two photon coherence \cite{ficek2010review}. As it has been mentioned in section \ref{model}, the two qubit system can be described in the computational basis given by $\{\ket{g_1}\ket{g_2}, \ket{g_1}\ket{e_2}, \ket{e_1}\ket{g_2}, \ket{e_1}\ket{e_2}\}$. Out of these, the state $\ket{g_1}\ket{g_2}$ is the ground state with no photons, $\{\ket{g_1}\ket{e_2}, \ket{e_1}\ket{g_2}\}$ are one photon excited states while the state $\ket{e_1}\ket{e_2}$ is two photon excited state. So, one photon coherence corresponds to coherence that is present between ground state and single photon ($\ket{g_1}\ket{g_2}\leftrightarrow \ket{g_1}\ket{e_2}, \ket{g_1}\ket{g_2}\leftrightarrow \ket{e_1}\ket{g_2}$), between two different single photon states ($\ket{e_1}\ket{g_2} \leftrightarrow \ket{g_1}\ket{e_2}$), and between single photon excited state and two photon excited state ($\ket{g_1}\ket{e_2}\leftrightarrow \ket{e_1}\ket{e_2}, \ket{e_1}\ket{g_2}\leftrightarrow \ket{e_1}\ket{e_2}$), while the two photon coherence represents the coherence between the ground state and two photon excited states ($\ket{g_1}\ket{g_2}\leftrightarrow \ket{e_1}\ket{e_2}$. The $X$-states corresponds to those states for which coherence between ground state and single photon states as well as coherence between single photon state and two photon coherence is zero. Further, the MNMSs and MEMSs formed by one photon and two photon coherence are distinguishable through the density matrix elements. Next, we elaborate upon the entanglement dynamics of MNMS and MEMS.

\subsection{Maximally nonlocal mixed states}
The maximally nonlocal mixed states \cite{batle2011nonlocality} forms a  subclass of $X$-states, and it produces the maximal violation of the Clauser–Horne–Shimony–Holt (CHSH) inequality \cite{clauser1969proposed}. This state is a Bell diagonal state  and the representation of MNMS for two photon coherence in the computational basis is
\begin{equation}
\rho^{MNMS}_{2} =
\begin{Bmatrix}
 \begin{pmatrix}
\frac{1}{2} & 0 & 0 & \frac{x}{2}\\
0 & 0 & 0 & 0\\
0 & 0 & 0 & 0\\
\frac{x}{2} & 0 & 0 & \frac{1}{2} 
\end{pmatrix}; 0 < x \leq 1
\end{Bmatrix},
\end{equation}
while the density matrix of MNMS for one photon coherence is written as
\begin{equation}
\rho^{MNMS}_{1} =
\begin{Bmatrix}
 \begin{pmatrix}
0 & 0 & 0 & 0\\
0 & \frac{1}{2} & \frac{x}{2} & 0\\
0 & \frac{x}{2} & \frac{1}{2} & 0\\
0 & 0 & 0 & 0 
\end{pmatrix}; 0 < x \leq 1
\end{Bmatrix}.
\end{equation}
For the range $0 < x \leq 1$, we see that MNMS remains in entangled form. For studying the entanglement dynamics, we consider this range  and analyze the sensitivity of the entanglement dynamics with respect to the initial amount of coherence. We explicitly consider a few values of $x$ and show the phenomena of ESD and ESB for both the two photon coherence and one photon coherence MNMS.

For studying the evolutionary dynamics, we use collective basis introduced in Section \ref{model}. In the collective basis, the matrix representation of two photon coherence MNMS evaluated at $r_{12} = \frac{\lambda}{6}, (\Gamma_{12} = 0.79\Gamma, \Omega_{12} = 1.12\Gamma$) for any time and for the {initial} coherence $x$  is computed as
\begin{equation}
\rho^{MNMS}_2 (t) =
%\begin{Bmatrix}
 \begin{pmatrix}
1 -a_{22}-a_{33}-a_{44}  & 0 & 0 & \frac{1}{2}e^{-t} x\\
 
0 & a_{22} & 0 & 0\\

0 & 0 & a_{33} & 0 \\

\frac{1}{2}e^{-t} x &0  & 0  & a_{44}
\end{pmatrix},  
%\end{Bmatrix}.
\end{equation}
where $a_{22}=4.3 e^{-2 t} (-1 + e^{0.21 t}), a_{33}=0.1 e^{-2 t} (-1 + e^{1.79 t}),$ and  $a_{44}=\frac{e^{-2 t}}{2}$. Similarly, the time dependent density matrix representation of the one photon coherence MNMS is computed as  
\begin{equation}
\rho^{MNMS}_1 (t) =
%\begin{Bmatrix}
 \begin{pmatrix}
1 - \frac{1}{2}e^{-0.21t}(1 - x) - \frac{1}{2} e^{-1.79 t} (1 + x) & 0 & 0 & 0\\
0 & \frac{1}{2} e^{-1.79 t} (1 + x) & 0 & 0\\

0 &0 & \frac{1}{2} e^{-0.21 t} (1 - x) & 0\\

0 & 0 & 0 & 0 
\end{pmatrix}.  
%\end{Bmatrix}.
\end{equation}
These matrices helps in describing the ESD and ESB for the MNMS two photon coherence and one photon coherence states. 

\begin{figure}[]
\centering
\includegraphics[width=0.85\linewidth]{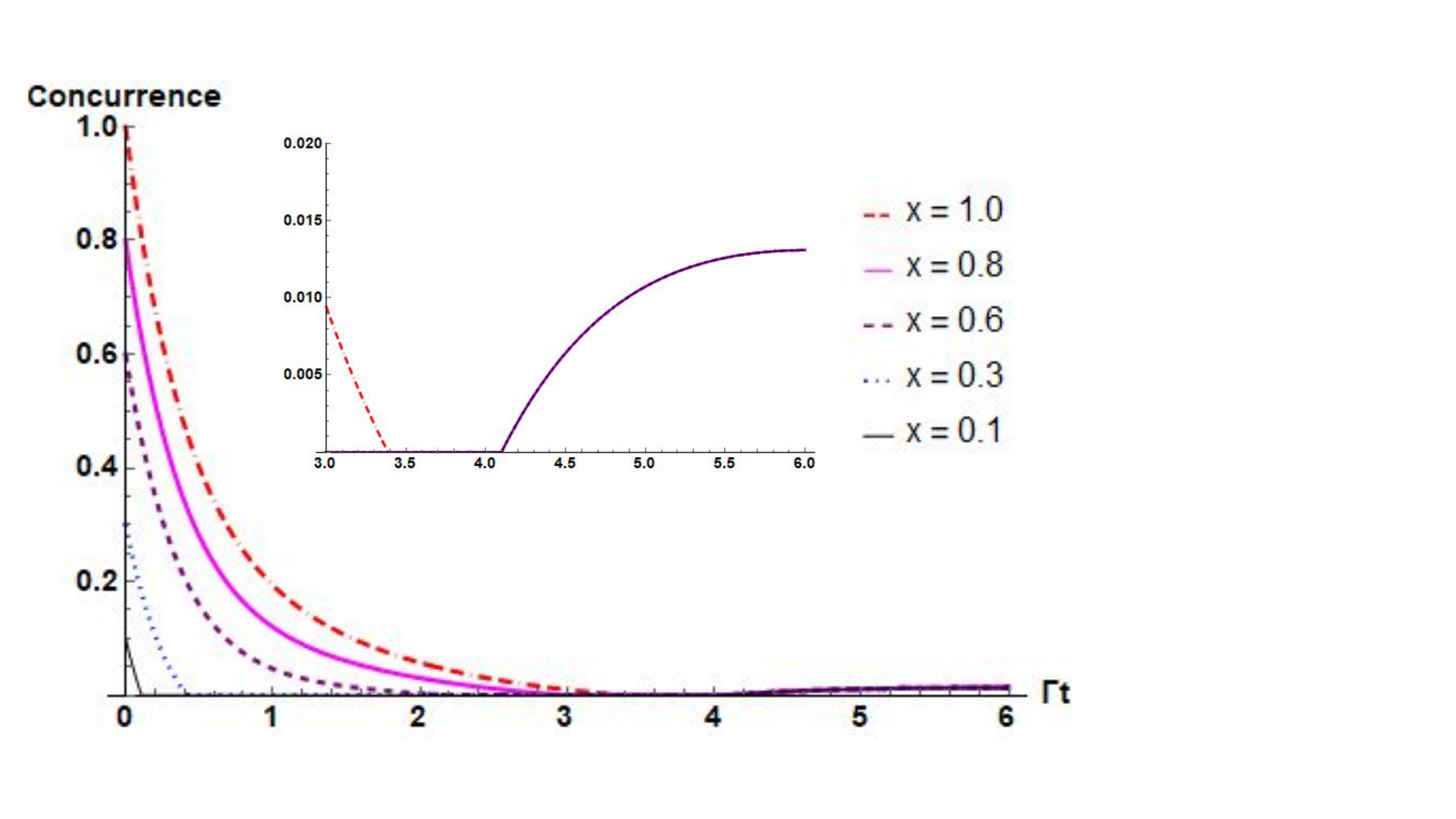}
\caption{(Color Online) Quantities plotted in this figure and all the other figures are dimensionless.
Concurrence as a function of dimensionless time parameter ($\Gamma t$) is plotted for the two photon coherence MNMSs for the different initial coherence ($x$) and the parameters $r_{12} = \frac{\lambda}{6}, (\Gamma_{12} = 0.79\Gamma, \Omega_{12} = 1.12\Gamma$). We have considered $x = 0.1, 0.3, 0.6, 0.8, 1.0 $ respectively, and plotted the evolutionary dynamics of MNMS given by $\rho_{2}^{MNMS}$ representing the entanglement sudden death and entanglement revival. The inset illustrates the revival of entanglement for all values of $x$. }
\label{figure-1}
\end{figure}

\begin{figure}[]
\centering
\includegraphics[width=0.67\linewidth]{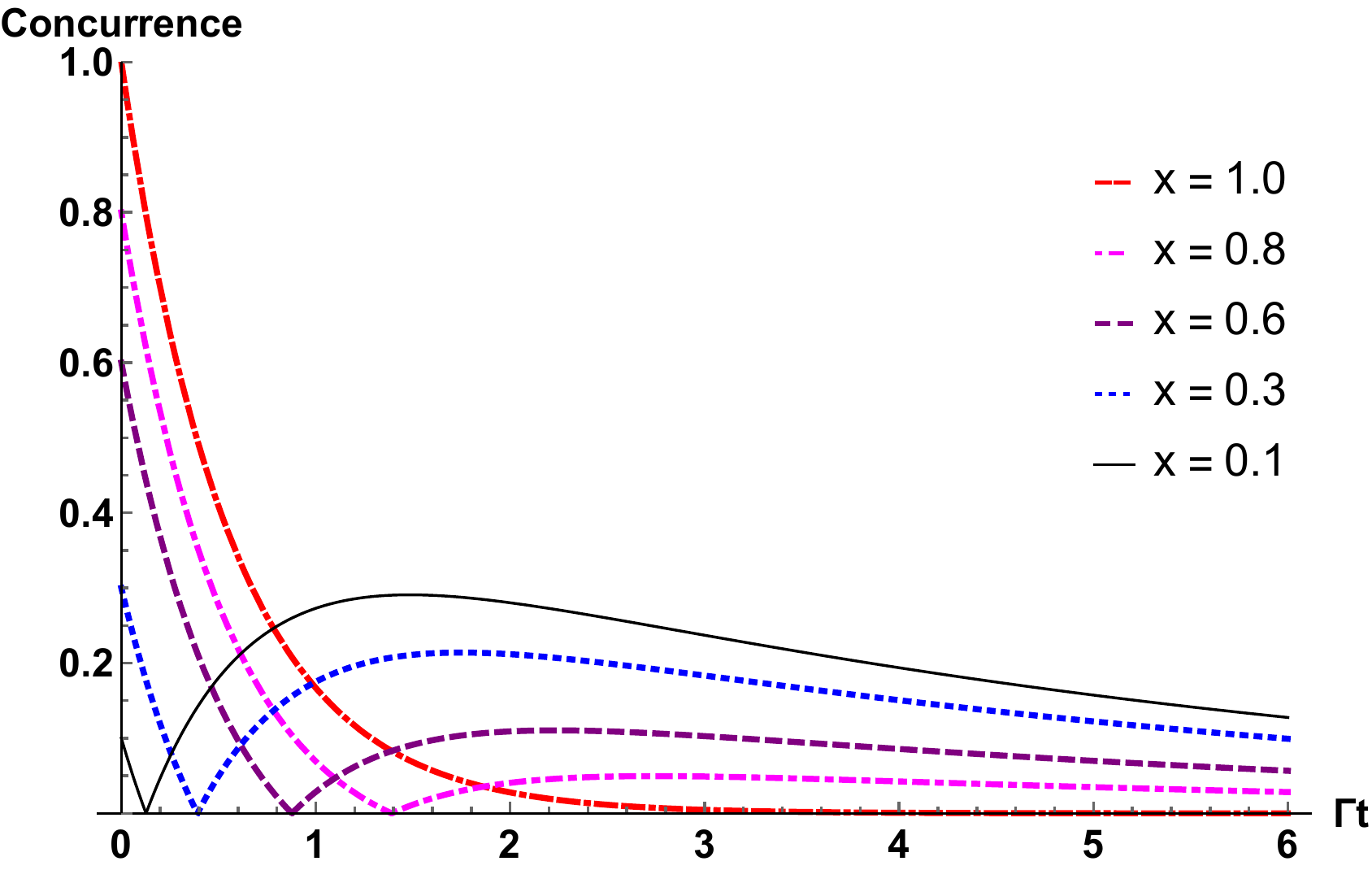}
\caption{(Color Online) Concurrence as a function of dimensionless time parameter ($\Gamma t$) is plotted for the one photon coherence MNMSsfor the different initial coherence ($x$) and the parameters $r_{12} = \frac{\lambda}{6}, (\Gamma_{12} = 0.79\Gamma, \Omega_{12} = 1.12\Gamma$). We have considered $x = 0.1, 0.3, 0.6, 0.8, 1.0 $ respectively, and plotted the evolutionary dynamics of MNMS given by $\rho_{1}^{MNMS}$. The plot represents the entanglement sudden death and entanglement revival phenomena for different values $x$.}
\label{figure-2}
\end{figure}

Figure \ref{figure-1} illustrates the entanglement dynamics for two photon coherence MNMS with concurrence (Eq. \ref{con_fun}) plotted as a function of dimensionless time parameter ($\Gamma t$) for different values of $x$ in the range $0 < x \leq 1$. The evolved dynamics is the result of interaction of two qubits with the vacuum reservoir. It can be followed from the figure that for all states (i.e, for all values $x$) the system exhibits ESD and ESB (cf. the plot given as inset in Fig. \ref{figure-1}). It can be seen that the entanglement decays to zero for different initial values of coherences (given by $x$ in figure) at different times, and for larger values of coherence the entanglement death occurs at a later time. Besides this, in a finite time the birth phenomenon is observed at the same time for all values of $x$, as can be seen from the subplot. Hence, it can be said that, the time at which entanglement decays is a sensitive function of the initial conditions given by the parameter $x$ while the revival time of entanglement is independent of $x$.  The inset in Fig. \ref{figure-1} illustrates the entanglement revival for all $x$ at the same time. This is due to the fact that, all the initial coherence terms decays and the population terms $a_{22}(t), a_{33}(t)$ becomes positive for collective bath at same time for all values of $x$, leading to entanglement birth at same time.

We also plot the entanglement dynamics of single photon coherence MNMS evolving under the vacuum bath in Fig. \ref{figure-2}. It depicts the collective decoherence of two qubits under the vacuum bath for different initial values of $x$ i.e., for different initial values of coherence.  We analysed the entanglement dynamics and compared it with that for the case of two photon coherence MNMSs. The time for ESD is dependent on the initial values of coherence (i.e. $x$) with states having lower value $x$ getting ESD at shorter time scales in comparison to that for the case of larger values of $x$.  Other than this, the plot also shows a very interesting behavior in which we observe instant revival of entanglement for all values of $x$. This is due to peculiar nature of concurrence function in which we can see that the threshold coherence term in the matrix $\rho_1^{MNMS}$ is zero and thus the entanglement instantly transfers to vacuum bath leading to the revival in the collective bath regime. It can be said that for the single photon coherence MNMS, entanglement birth occurs rapidly and they revive with higher amplitudes in comparison to two photon coherence MNMS. Thus, these states are less vulnerable to the effects of vacuum bath and continue to maintain entanglement over time, and thus can play an important role in the implementation of quantum algorithms and protocols.

\subsection{Maximally entangled mixed states}
Now, we will deal with the other class of $X$-states known as MEMS \cite{munro2001maximizing}. These
states are a generalization of the class of Bell states to mixed states and are known
to have the highest degree of entanglement for a given purity of the state. The MEMSs for two photon coherence can be expressed as 
\begin{equation}
\rho^{MEMS}_{2} =
\begin{Bmatrix}
 \begin{pmatrix}
g(x) & 0 & 0 & \frac{x}{2}\\
0 & 1-2g(x) & 0 & 0\\
0 & 0 & 0 & 0\\
\frac{x}{2} & 0 & 0 & g(x) 
\end{pmatrix};  g(x)= \frac{1}{3}\,  {\rm{if}} \; x < \frac{2}{3} \; {\rm{and}} \, g(x)= \frac{x}{2}\,  {\rm{if}}\;  \frac{2}{3} \leq x \leq 1 
\end{Bmatrix},
\end{equation}
while the MEMSs for one photon coherence can be expressed as 
\begin{equation}
\rho^{MEMS}_{2} =
\begin{Bmatrix}
 \begin{pmatrix}
1-2g(x) & 0 & 0 & 0\\
0 & g(x) & \frac{x}{2} & 0\\
0 & \frac{x}{2} & g(x) & 0\\
0 & 0 & 0 & 0 
\end{pmatrix};  g(x)= \frac{1}{3}\,  {\rm{if}} \; x < \frac{2}{3} \; {\rm{and}} \, g(x)= \frac{x}{2}\,  {\rm{if}}\;  \frac{2}{3} \leq x \leq 1 
\end{Bmatrix}.
\end{equation}
Here, we can see that the coherence $\frac{x}{2}$ for single photon coherence state corresponds to coherence $\ket{e_1}\ket{g_2} \leftrightarrow \ket{g_1}\ket{e_2}$ while two photon coherence state corresponds to coherence $\ket{g_1}\ket{g_2} \leftrightarrow \ket{e_1}\ket{e_2}$. Similar to the MNMSs, we will analyze the entanglement dynamics of the both two photon and one photon coherence MEMSs.

\begin{figure}[h!]
\centering
\includegraphics[width=0.7\linewidth]{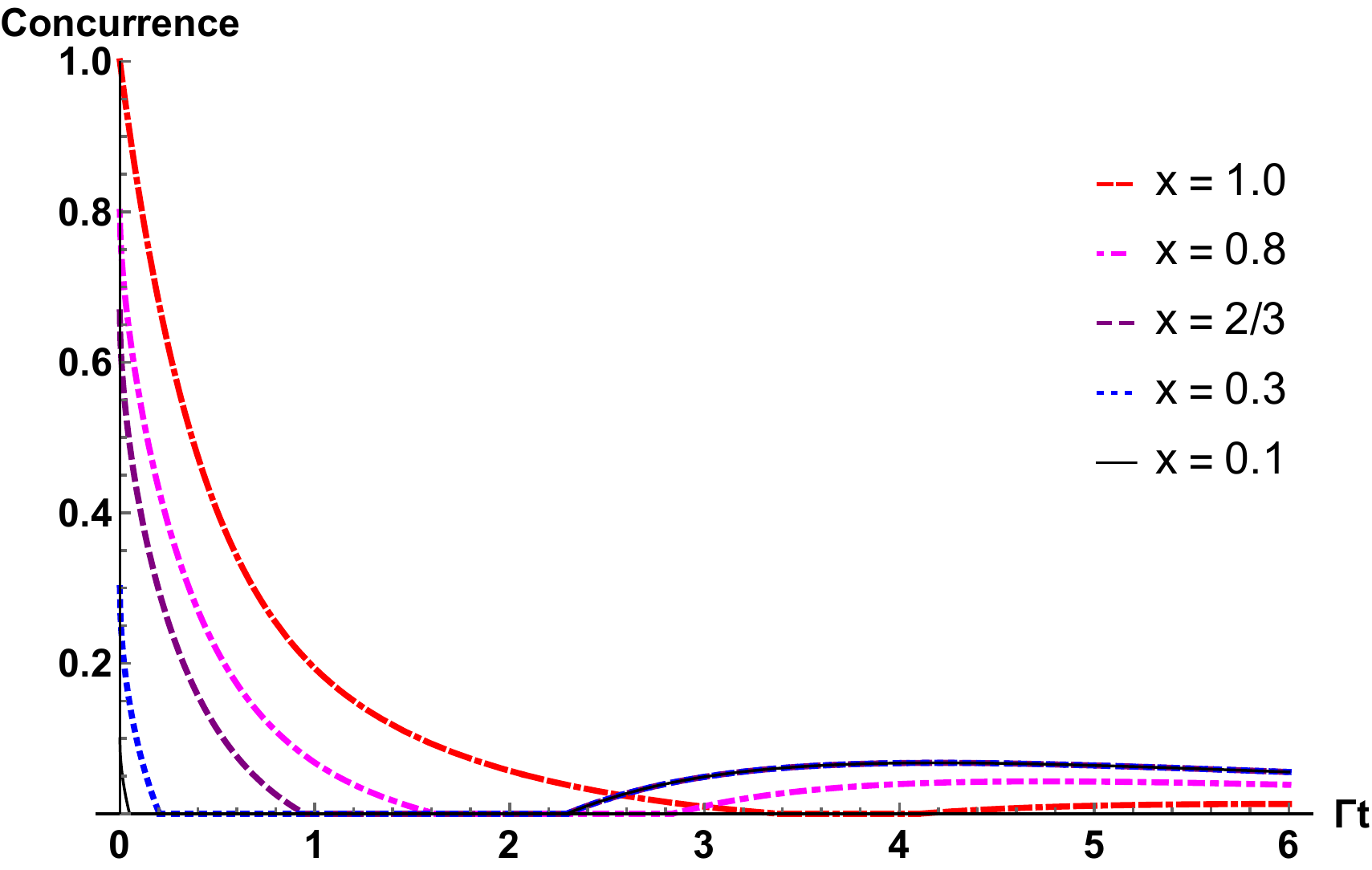}
\caption{(Color Online) Concurrence as a function of dimensionless time parameter ($\Gamma t$) is plotted for the two photon coherence MEMSs for the different values of initial coherence $x$ and the parameters $r_{12} = \frac{\lambda}{6}, (\Gamma_{12} = 0.79\Gamma, \Omega_{12} = 1.12\Gamma$). We have considered $x = 0.1, 0.3, 2/3, 0.8, 1.0 $ respectively, and plotted the evolutionary dynamics of MEMS. The plot represents the entanglement sudden death and entanglement revival phenomena for all the coherent states of $x$.}
\label{figure:3}
\end{figure}
\begin{figure}[h!]
\centering
\includegraphics[width=0.85\linewidth]{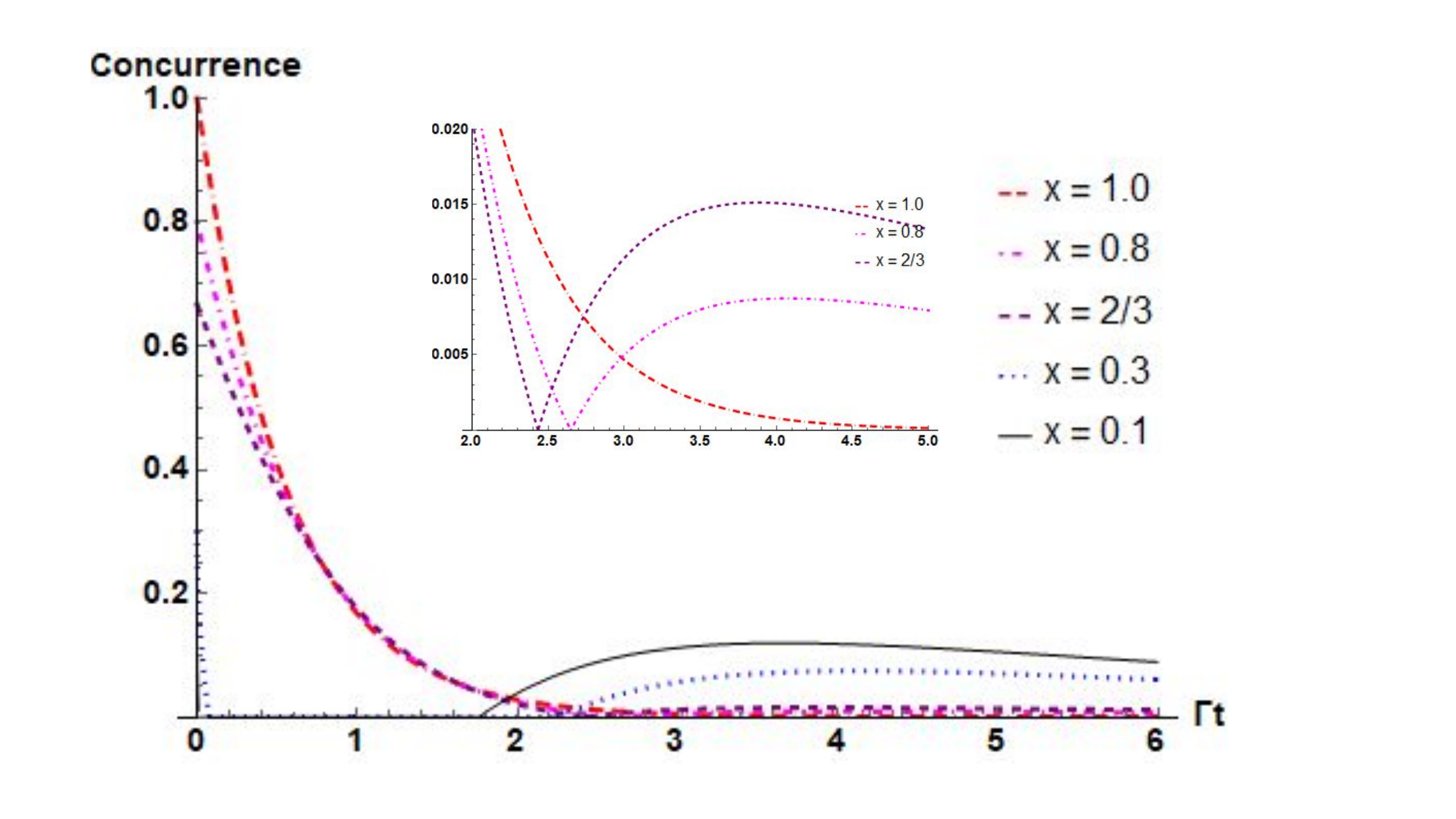}
\caption{(Color Online) 
Concurrence as a function of dimensionless time parameter ($\Gamma t$) is plotted for the one photon coherence MEMSs for the different values of initial coherence $x$ and the parameters $r_{12} = \frac{\lambda}{6}, (\Gamma_{12} = 0.79\Gamma, \Omega_{12} = 1.12\Gamma$). We have considered $x = 0.1, 0.3, 2/3, 0.8, 1.0 $ respectively, and plotted the evolutionary dynamics given by $\rho_{1}^{MEMS}$ in the range of $0 < x \leq 1$. The plot represents the entanglement sudden death and entanglement revival. The inset here illustrates the revival of entanglement for $x = 2/3, 0.8, 1.0$. }
\label{figure:4}
\end{figure}

We may now discuss the entanglement dynamics for two photon coherence and one photon coherence MEMSs when the two qubit system with given initial conditions evolves under vacuum bath. The same is illustrated in Figs. \ref{figure:3} and  \ref{figure:4}. It can be  observed from the figures that for larger values of $x$, ESD occurs at a relatively higher value of time. We can see from Fig. \ref{figure:3} that for the case of  two photon coherence MEMSs with initial coherence lying in the range $0 < x \leq \frac{2}{3}$, the entanglement rebirth occurs at the same time and have same amount of entanglement amplitude (as measured via coherence) for all values of $x$. For the case of initial coherence  $x > \frac{2}{3}$,  rebirth of entanglement is observed at later time intervals. Further, the amplitude of concurrence after rebirth for $x > \frac{2}{3}$ is smaller than that for the case of $0 < x \leq \frac{2}{3}$. The entanglement dynamics for the case of  one photon coherence MEMS is shown in Fig. \ref{figure:4}. We can see that the entanglement rebirth occurs sooner for smaller values of initial coherence ($x$). Further, for the larger values of $x$, the rebirth occurs with a reduced amplitude. Another point to be noted is that for the same value of initial coherence, ESD for one photon coherence happens at a later time in comparison to that for the case of two photon coherence MEMS. Hence, the one photon coherence states can be found to be advantageous over two photon coherence states.

\section{MNMS and MEMS states under noisy channels} \label{noise}
A quantum system evolving under the effect of ambient environment \cite{caruso2014quantum} can be described by completely positive trace preserving (CPTP) map from an initial state to final state. The operator sum representation of the final state is written as \cite{kraus1971general} 
\begin{equation}
    \rho(t) = \sum_{i} E_{i}(t)\rho(0) E_{i}^{\dag}(t),
\end{equation} here $E_i's$ are the Kraus operators with $\sum_{i}E_{i}^{\dag}E_{i} = I$ for all $t$.\\
In this section, we will discuss the effect of different noisy channels like phase damping \cite{kraus1983states, yu2005evolution}, amplitude damping \cite{kraus1983states, yu2005evolution, srikanth2008squeezed}, and subsequently the random telegraph noise \cite{naikoo2019facets} for the single photon and two photon coherence MNMS and MEMS. 

\subsection{Phase damping channel}
The phase damping channel \cite{nielsen2000quantum} corresponds to the decoherence in the system, in which there is no energy loss while the phase dies out. For the evolution of the system, the Kraus operators for a single qubit system realising the phase damping channel are given by
\begin{equation}
    E_{0}^{PD} = \begin{pmatrix}
    1 & 0\\
    0 & \gamma(t)
    \end{pmatrix},\\ \space
    E_{1}^{PD} = \begin{pmatrix}
    0 & 0\\
    0 & \omega(t)
    \end{pmatrix},
\end{equation}
where the time-dependent matrix elements are given as $\gamma(t) = \exp{(-\Gamma_{PD}t/2)}$, and $\omega(t) = \sqrt{1-\gamma^{2}(t)}$ with $\Gamma_{PD}$ representing the phase damping rate of the qubit.

\begin{figure}[h!]
    \centering
    {{\includegraphics[width=8.2cm]{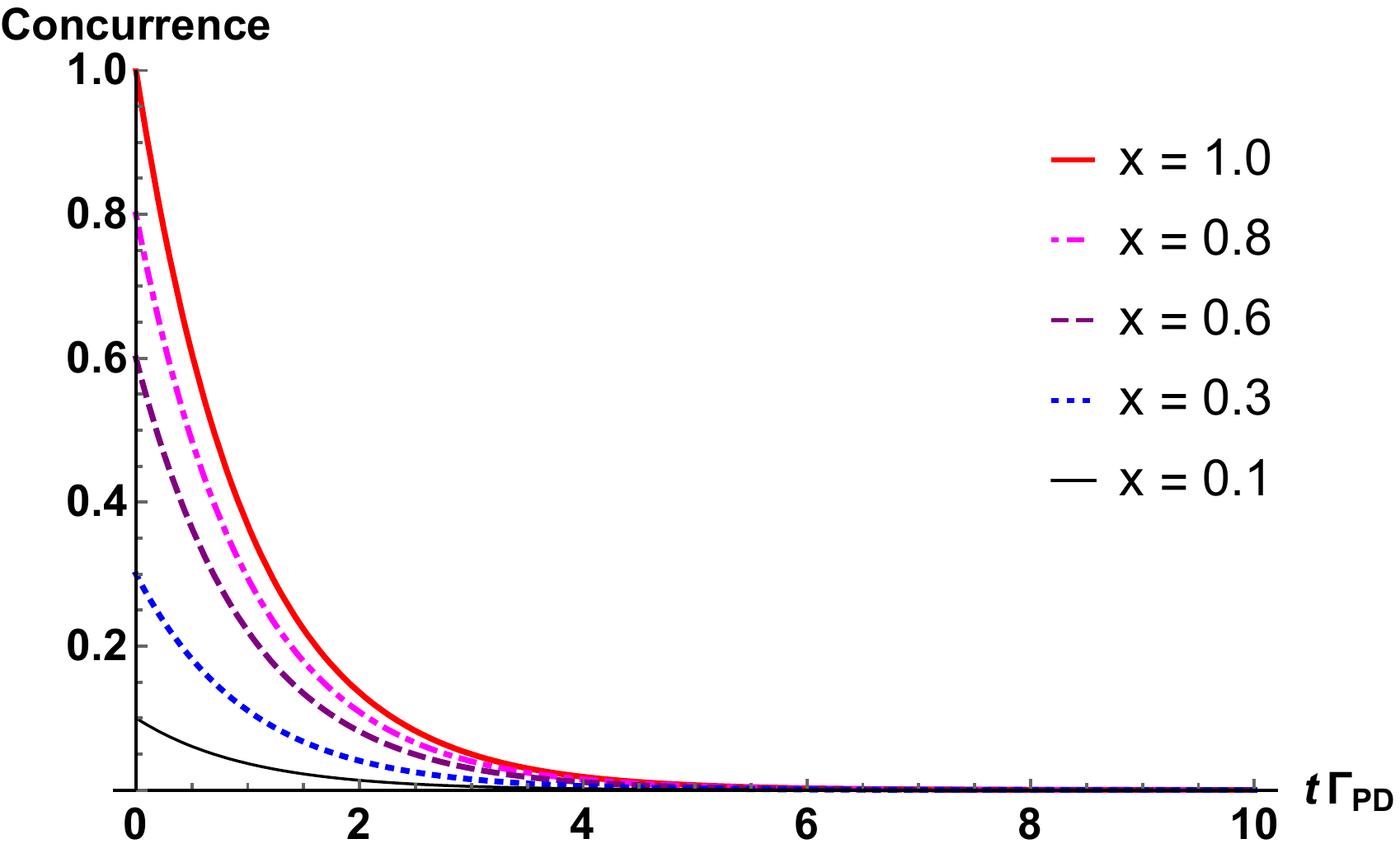}}}
    \qquad
   {{\includegraphics[width=8.2cm]{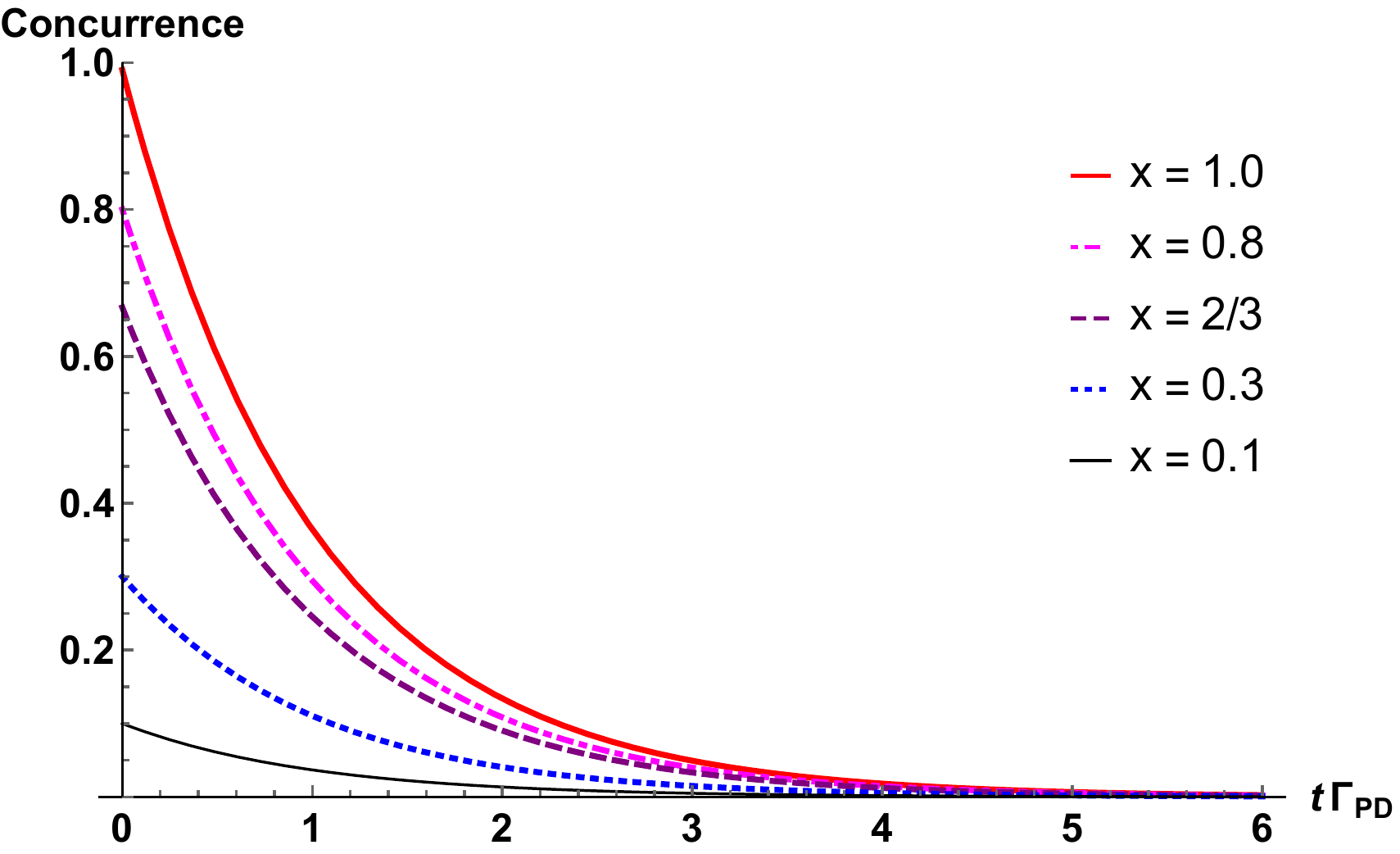}}}
   \centerline{ \small (a) \hspace{.4\hsize} (b) }
    \caption{(Color online) The plots shows the evolutionary dynamics when $C(t)$ is plotted against dimensionless time parameter ($\Gamma_{PD}t$) for (a) MNMS two photon coherence (b) MEMS two photon coherence evolving under the dephasing channel. Similar behavior is observed for one photon coherence MNMS ad MEMS. }
    \label{fig-5}
\end{figure}

In this section, we have studied evolutionary dynamics of master equation using the phase damping Kraus operators for the two photon coherence and one photon coherence of MNMS and MEMS. Under phasing damping, the density matrix representation for the two photon coherence  MNMS at any time is computed as  
\begin{equation}
\rho^{MNMS}_2 (t) =
%\begin{Bmatrix}
 \begin{pmatrix}
\frac{1}{2} & 0 & 0 & \frac{x}{2}e^{-\Gamma_{PD}t}\\
0 & 0 & 0 & 0\\
0 & 0 & 0 & 0\\
\frac{x }{2}e^{-\Gamma_{PD}t} & 0 & 0 & \frac{1}{2}
\end{pmatrix},  
%\end{Bmatrix}.
\end{equation}
while that for the case of single photon coherence MNMS is computed as 
\begin{equation}
\rho^{MNMS}_1 (t) =
%\begin{Bmatrix}
 \begin{pmatrix}
0 & 0 & 0 & 0\\
0 & \frac{1}{2} & \frac{x }{2}e^{-\Gamma_{PD}t} & 0\\

0 & \frac{x}{2}e^{-\Gamma_{PD}t} & \frac{1}{2} & 0\\

0 & 0 & 0 & 0 
\end{pmatrix}.  
%\end{Bmatrix}.
\end{equation}
 Fig. \ref{fig-5} (a) shows the  time evolution of concurrence for two photon coherence  MNMSs. Applying the formula of concurrence, both for the two photon as well as single photon coherence MNMSs, the threshold coherence is zero while coherence term $\frac{x}{2}e^{-2\Gamma_{PD}t}$ goes to zero asymptotically depending on the initial values of coherence $x$. So, there is no difference for the entanglement dynamics of two photon coherence and single photon coherence MNMSs under phase damping. Similar, is the entanglement dynamics for the case of MEMSs as can be seen from Fig. \ref{fig-5} (b) .

\subsection{Amplitude damping channel}
The amplitude damping channel encaptures the idea of energy dissipation from a system. For example in two level atom model, due to spontaneous emission of photons, the system decays in time. For our purpose, when the two qubit system is interacting with vacuum bath given by Eq. \ref{Eqn.2}, we employ the noise operators for the evolution and study decoherence. The corresponding Kraus operators for the amplitude damping noise channel for single qubit are given by
\begin{equation}\label{Eqgn-24}
    E_{0}^{AD} = \begin{pmatrix}
    \gamma(t) & 0\\
    0 & 1
    \end{pmatrix},\\ \space
    E_{1}^{AD} = \begin{pmatrix}
    0 & 0\\
    \omega(t) & 0
    \end{pmatrix},
\end{equation} 
where $\gamma(t) = \exp{(-\Gamma_{AD}t/2)}$, and $\omega(t) = \sqrt{1-\gamma^{2}(t)}$ with $\Gamma_{AD}$ representing the amplitude damping rate acting on individual qubits. \\
Under the amplitude damping channel, the density matrix representation for the two photon coherence MNMS at any time is computed as  
\begin{equation}
\rho^{MNMS}_2 (t) =
%\begin{Bmatrix}
 \begin{pmatrix}
\frac{1}{2}e^{-2\Gamma_{AD}t} & 0 & 0 & \frac{x}{2}e^{-\Gamma_{AD}t}\\
0 & \frac{1}{2}e^{-\Gamma_{AD}t}(1-e^{-\Gamma_{AD}t}) & 0 & 0\\
0 & 0 &  \frac{1}{2}e^{-\Gamma_{AD}t}(1-e^{-\Gamma_{AD}t}) & 0\\
\frac{x }{2}e^{-\Gamma_{AD}t} & 0 & 0 & \frac{1}{2} + \frac{1}{2}(1 - e^{-\Gamma_{PD}t})^2
\end{pmatrix},  
%\end{Bmatrix}.
\end{equation}
while that for the case of single photon coherence of MNMS is computed as 
\begin{equation}
\rho^{MNMS}_1 (t) =
%\begin{Bmatrix}
 \begin{pmatrix}
0 & 0 & 0 & 0\\
0 & \frac{1}{2}e^{-\Gamma_{AD}t} & \frac{x}{2}e^{-\Gamma_{AD}t} & 0\\
0 & \frac{x}{2}e^{-\Gamma_{AD}t} &  \frac{1}{2}e^{-\Gamma_{AD}t} & 0\\
0 & 0 & 0 & 1-e^{-\Gamma_{AD}t}
\end{pmatrix}.  
%\end{Bmatrix}.
\end{equation}

\begin{figure}[h!]
    \centering
    {{\includegraphics[width=8.2cm]{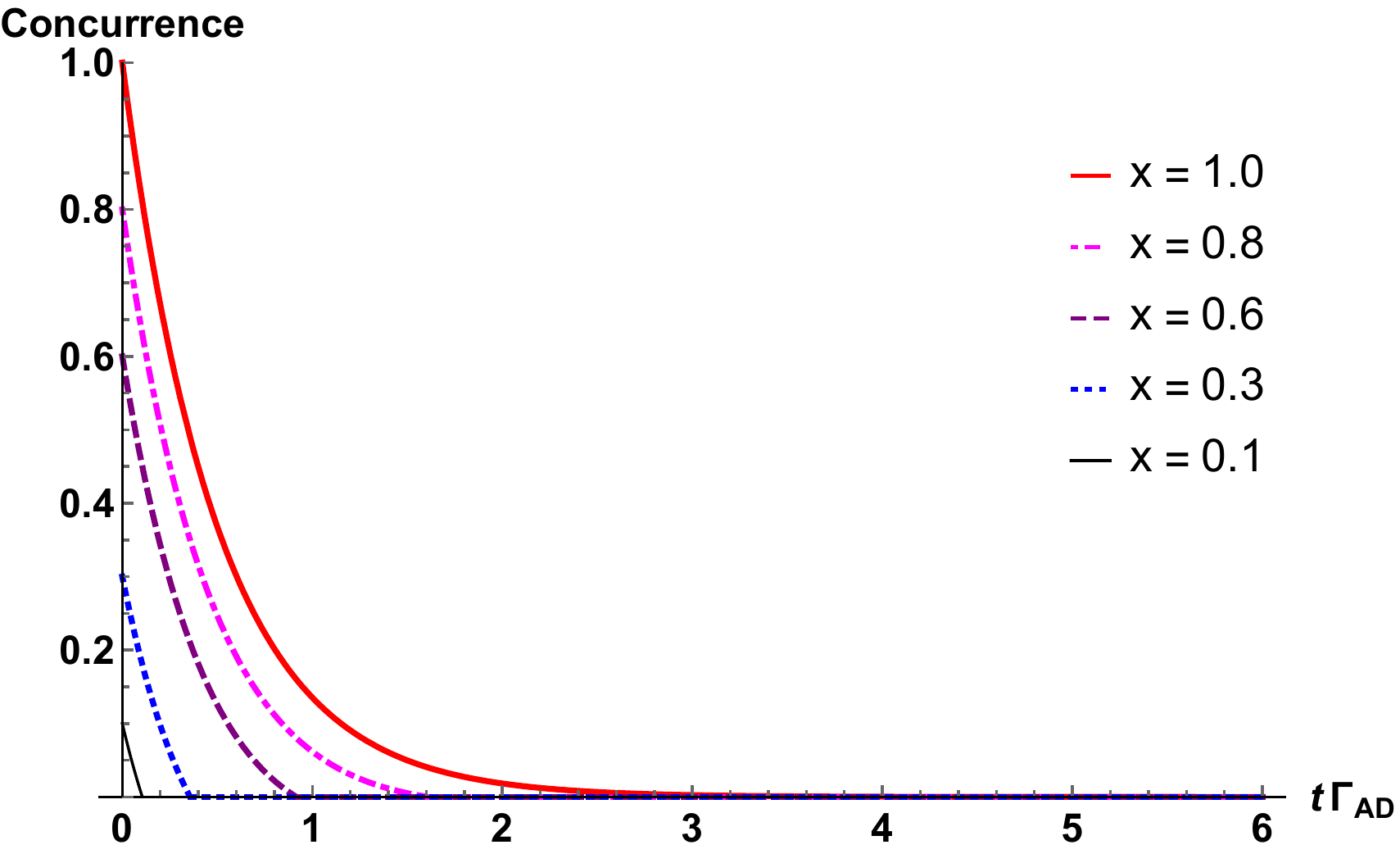}}}
    \label{a}
    \qquad
   {{\includegraphics[width=8.2cm]{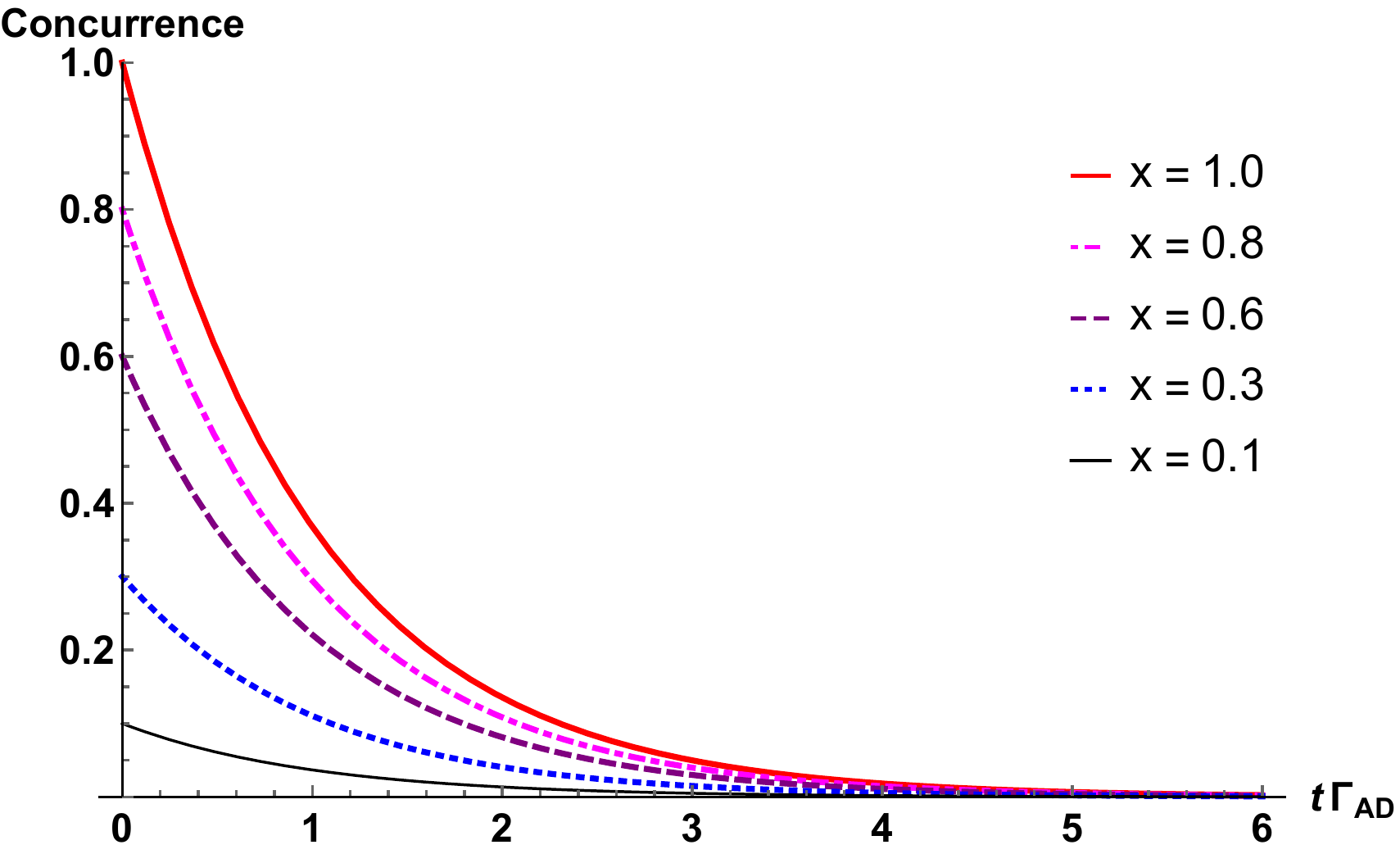}}}
   \centerline{ \small (a) \hspace{.4\hsize} (b) }
    \caption{(Color online) The plots show the evolutionary dynamics when $C(t)$ is plotted against dimensionless time parameter ($\Gamma_{AD}t$) for the (a) two photon coherence MNMS (b) one photon coherence MNMS evolved under amplitude damping channel. ESD is observed for two photon coherence states while single photon coherence states decay asymptotically. }
    \label{figure-7}
\end{figure} 

\begin{figure}[h!]
    \centering
    {{\includegraphics[width=8.2cm]{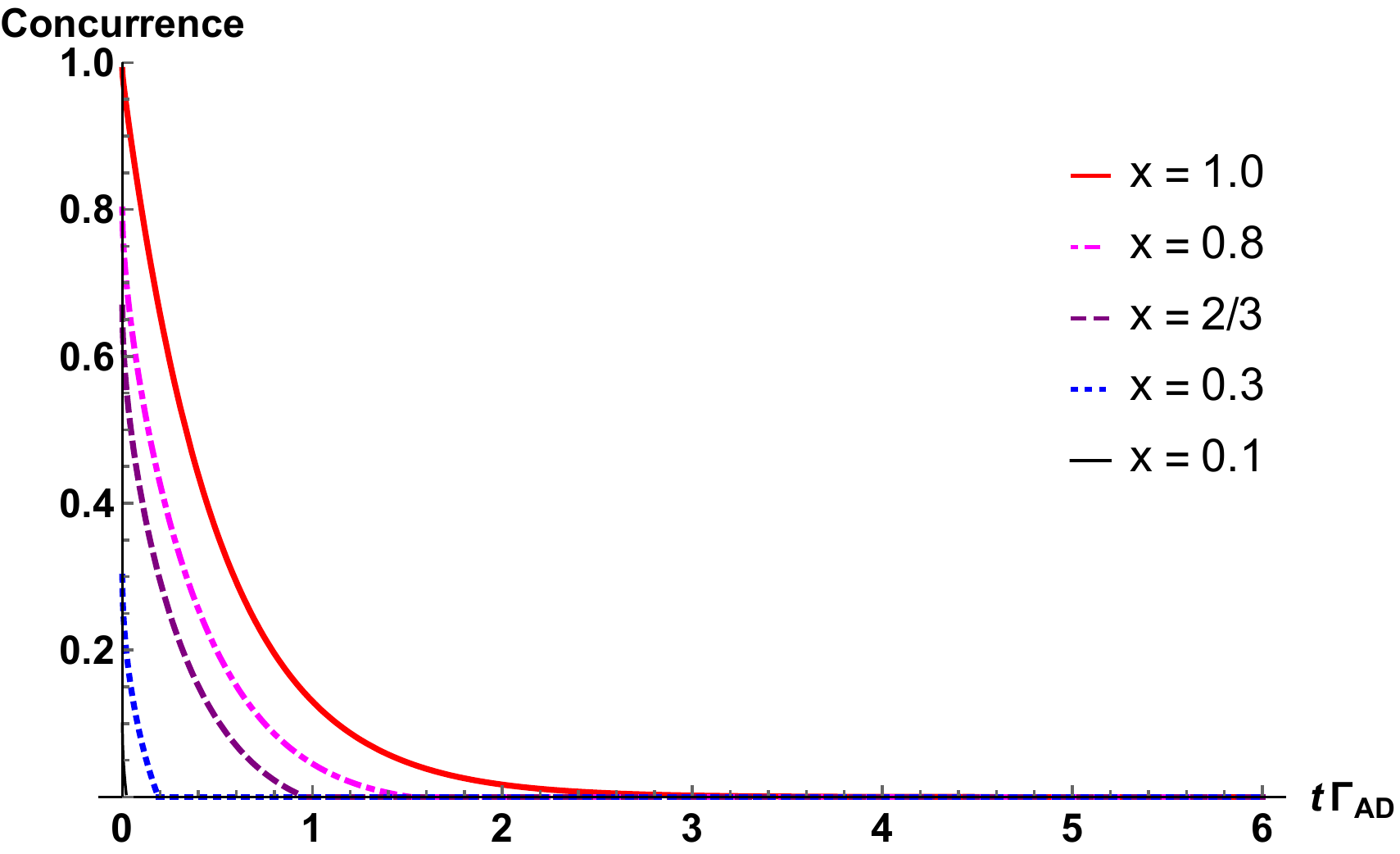}}}
    \qquad
   {{\includegraphics[width=8.2cm]{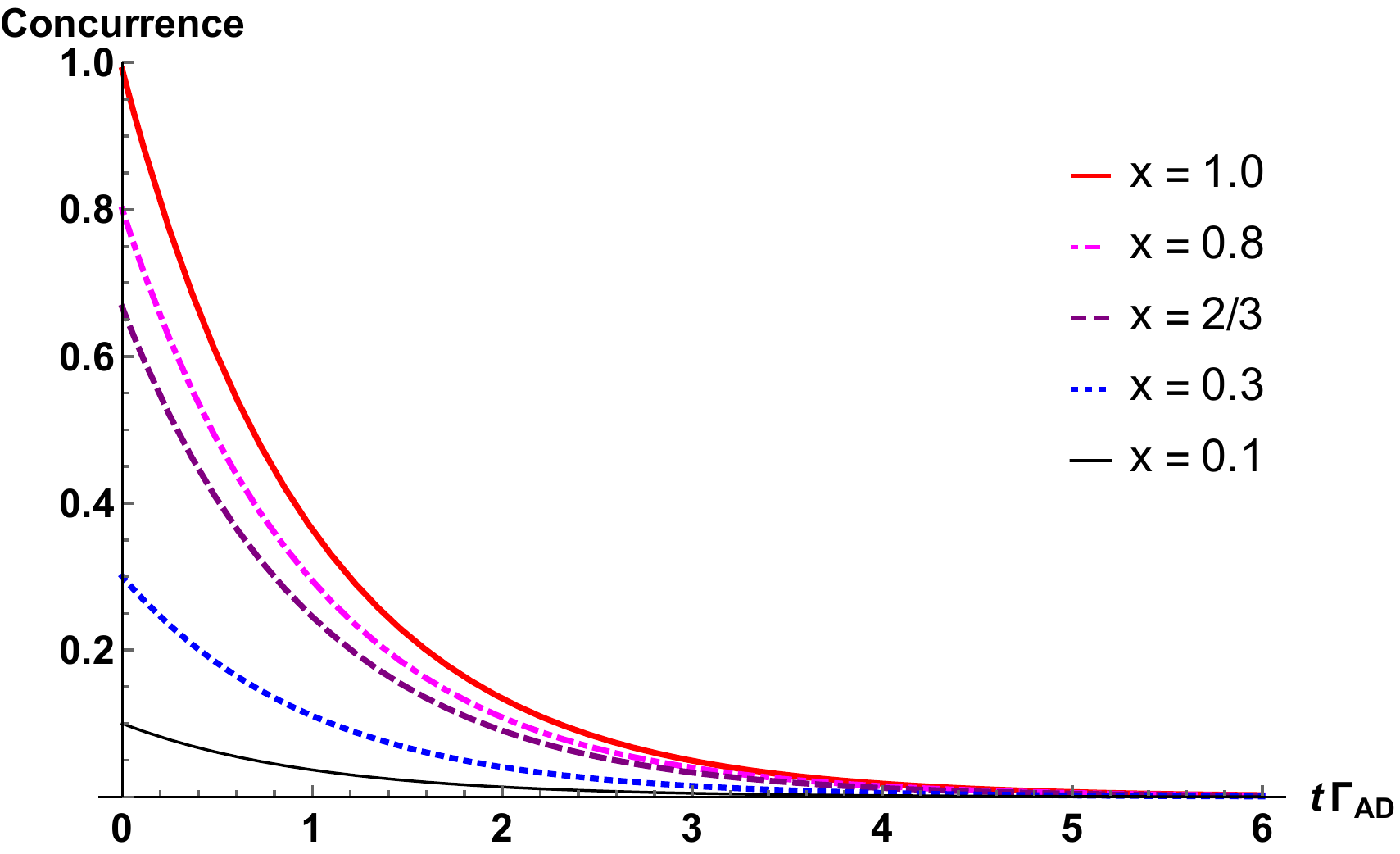}}}
   \centerline{ \small (a) \hspace{.4\hsize} (b) }
    \caption{(Color online) The plots show the evolutionary dynamics when $C(t)$ is plotted against dimensionless time parameter ($\Gamma_{AD}t$) for the (a) two photon coherence MEMS (b) one photon coherence MEMS evolved under amplitude damping channel.  ESD is observed for two photon coherence states while single photon coherence states decay asymptotically.}
    \label{fig-8}
\end{figure}

During the evolution of two photon coherence MNMS under the amplitude damping channel, we see that at any time, $\rho^{MNMS}_2 (t)$ depends on all the population terms as well as on $\rho_{14}(t)$ and the coherence term is given as $ \frac{x}{2}e^{-\Gamma_{AD}t}$. Fig. \ref{figure-7}(a) shows that under the amplitude damping channel the entanglement sudden death occurs for all the two photon coherence states of $x$.  The Figure \ref{figure-7}(b) shows the entanglement dynamics of one photon coherence state MNMS, Here, the threshold coherence is zero while the coherence term $\rho_{23}(t)$ varies as $ \frac{x}{2}e^{-\Gamma_{AD}t}$. So, we can see that   for all the initial values of coherence for one photon coherence MNMS, the entanglement decays asymptotically with no ESD. Hence, single photon coherence MNMS are more useful for practical applications in comparison to their two photon coherence counterpart. Similar is the entanglement dynamics for the case of two photon coherence and one photon coherence of MEMSs as is seen from Fig. \ref{fig-8}.

\subsection{Random telegraph noise}
The random telegraph noise \cite{kumar2018non} describe the dynamics when the initial state is exposed to a bi-fluctuating classical noise.  The corresponding Kraus operators representing random telegraph noise are given by 

\begin{equation} \label{Eqn-13}
    E_{0}^{RTN} = 
    \sqrt{\frac{1+\Delta(t)}{2}}\begin{pmatrix} 1 & 0\\
    0 & 1
    \end{pmatrix},\\ \space
    E_{1}^{RTN} = \sqrt{\frac{1-\Delta(t)}{2}}\begin{pmatrix} 1 & 0\\
    0 & -1
    \end{pmatrix},
\end{equation} where the time-dependent function $\Delta(t)$ is given as\\ $\Delta(t) = \exp{(-\gamma_{RTN} t)}\Bigg[\cos{\Big( \gamma_{RTN} t\sqrt{\big(\frac{2b}{\gamma_{RTN}}\big)^{2}-1 }} \Big) + \frac{\sin{\Big( \gamma_{RTN} t\sqrt{\big(\frac{2b}{\gamma_{RTN}}\big)^{2}-1 }} \Big)}{\sqrt{\big(\frac{2b}{\gamma_{RTN}}\big)^{2}-1 }} \Bigg]$. Here $b$ is the strength that quantifies system-environment coupling, $\gamma_{RTN}$ is the fluctuation rate of RTN. The ratio of $2b$ and $\gamma_{RTN}$ determines the regime with Markovian character if $(2b)^2 \leq \gamma_{RTN}^2 $ and non-markovian otherwise.
\begin{figure}[h!]
    \centering
    {{\includegraphics[width=8.2cm]{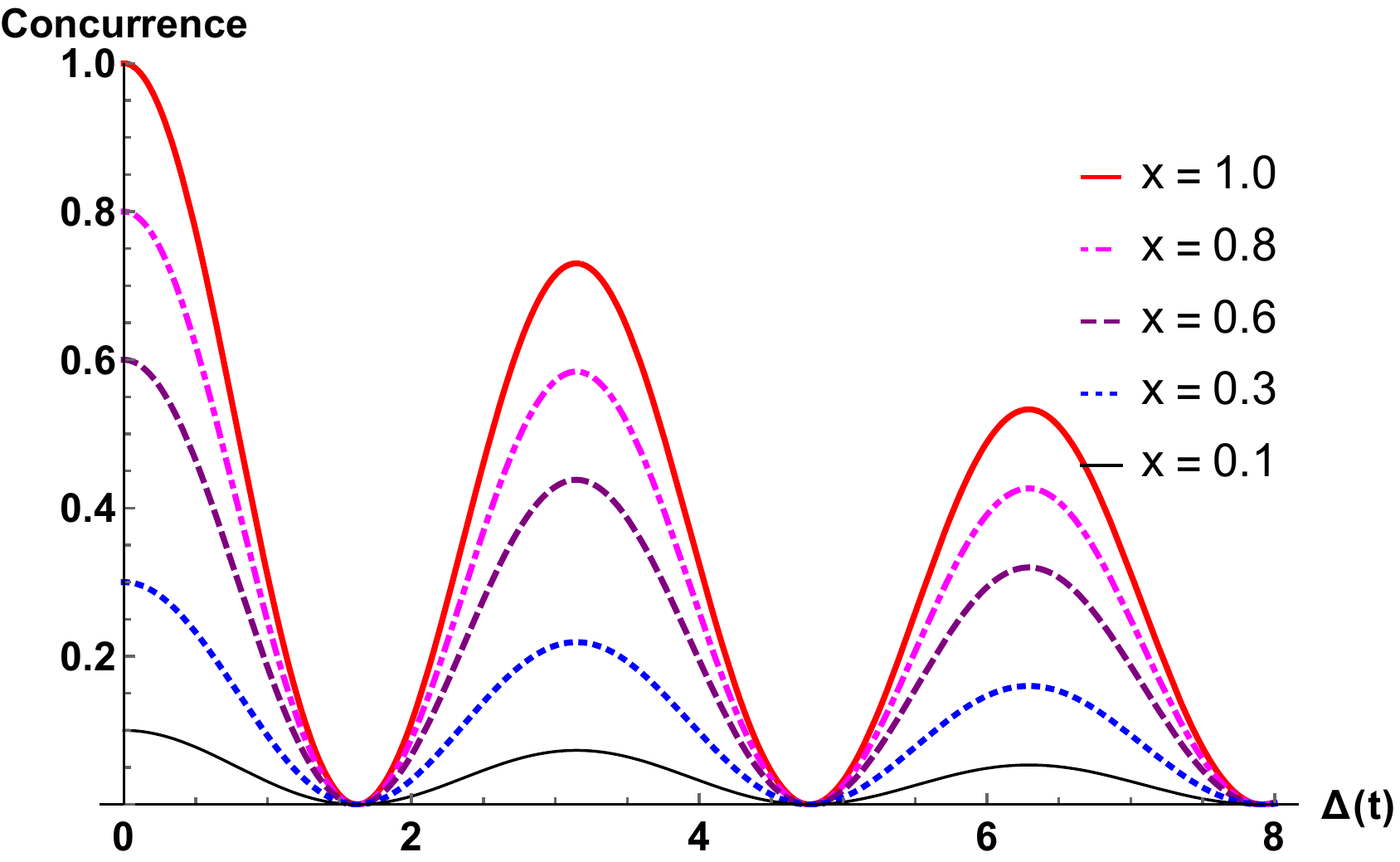}}}
    \qquad
   {{\includegraphics[width=8.2cm]{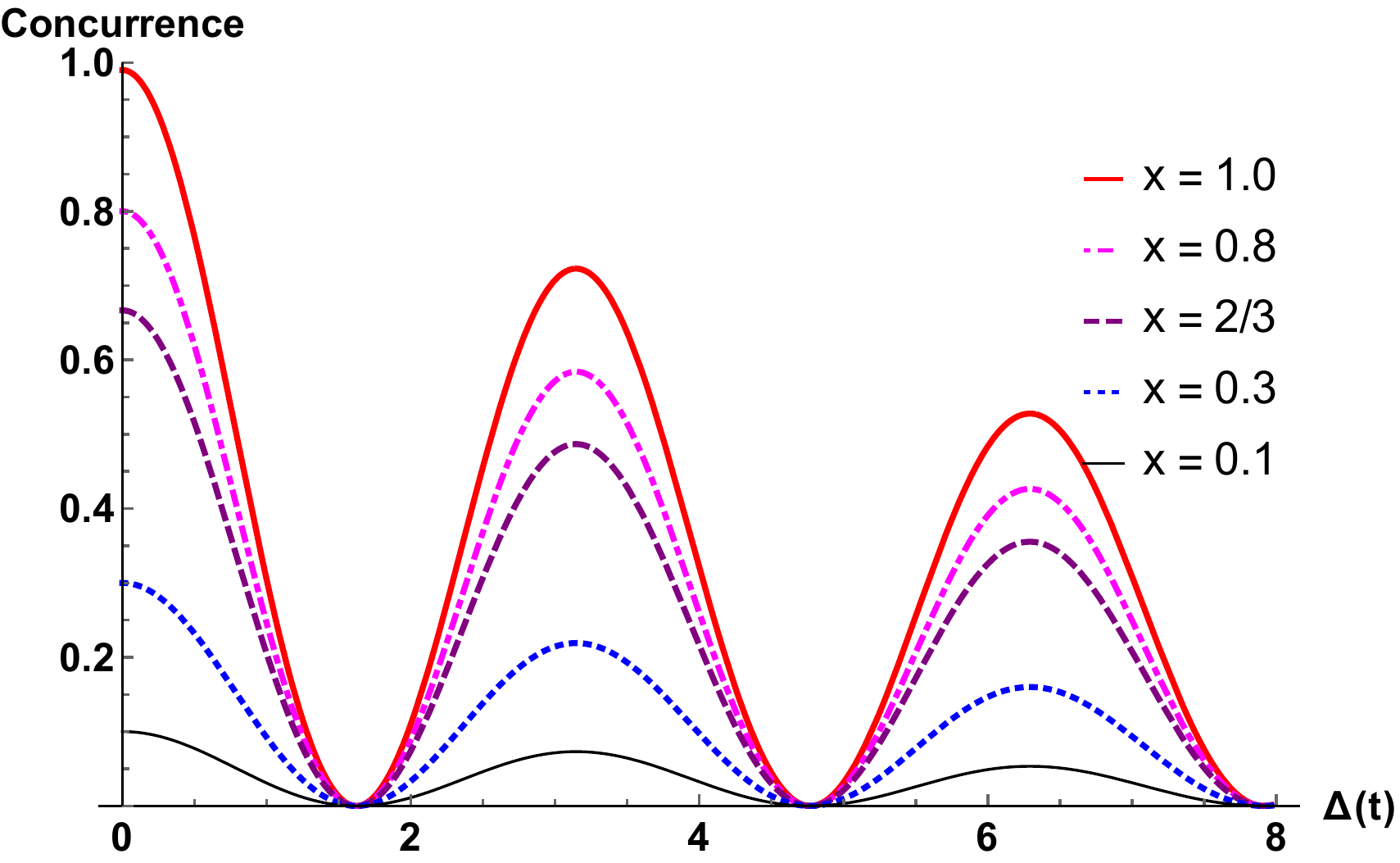}}}
   \centerline{ \small (a) \hspace{.4\hsize} (b) }
    \caption{The plots describe the evolutionary dynamics when the (a )MNMS (b) MEMS  (for the two photon coherence  states) evolved under RTN. It shows the decoherence and the entanglement revival for all $x$ in the range $0 < x \leq 1$ both the cases.}
    \label{figure-9}
\end{figure}
The MNMSs and MEMSs evolutionary dynamical structure of density matrix can be obtained by Kraus operators in \ref{Eqn-13}.

The dynamical behavior of these states for the two photon coherence under the action of RTN is illustrated in Fig. \ref{figure-9} in the non-Markovian regime. It can be noted from our results that in the non-Markovian regime, the function $\Delta(t)$ has an oscillatory nature due to the presence of the trigonometric functions in the expression \ref{Eqn-13}. Whereas in the Markovian regime, we see decay for all coherence states due to hyperbolic functions. Hence, it can be concluded that the coherence of the MNMSs and MEMSs decay monotonously in the Markovian regime, but in case of non-markovian regime it shows revival of entanglement of all values of $x$ (i.e., for all values of initial coherence). It can also be noted that with time, the amplitude of these states reduces leading to asymptotic decay. The similar behavior is noticed for both single photon coherence and two photon coherence MNMS and MEMS.

\section{Conclusion} \label{conclusion}

We have studied the open system quantum entanglement dynamics of $X$-states with a special emphasis on MNMS and MEMS by considering a model of two spatially separated two level atoms (two qubits) coupled to a common vacuum bath. Due to the effect of environment, the entanglement between the two qubits undergo environmental decoherence, and in many situations we observed the phenomena of ESD and  ESB. For both MNMSs and MEMSs, we tried to compare and contrast the entanglement dynamics observed for single photon coherence and two photon coherence states. When MNMSs and MEMSs were interacting with an environment consisting of vacuum bath, then  the phenomena of ESD and ESB were observed, but the characteristics of ESD and ESB were different for MNMS and MEMS. Even for same class of states (MNMS or MEMS), the two photon coherence and single photon coherence states were found to behave differently with respect to entanglement dynamics. It was observed that for both MNMS and MEMS, the greater amount of initial coherence delays the sudden death of entanglement. Further, it was observed that ESB happens earlier for the case of single photon coherence states in comparison to the two photon coherence states and for many situations the entanglement was observed to revive almost instantaneously for the case of single photon states. Further, entanglement dynamics of MNMS and MEMS were investigated under the presence of different noises namely phase damping, amplitude damping and RTN noise. We saw that under the presence of phase damping noise, entanglement for both MNMS and MEMS decays asymptotically with no ESD. However, for the amplitude damping noise, ESD was observed only for the case of two photon coherence MNMSs and MEMSs while the entanglement for single photon coherence states decaying only asymptotically. Further, no revivals of entanglement were observed. When the qubits were subjected to non-Markovian RTN noise, then  both the single photon coherence and two photon coherence MNMS and MEMS behaved identically with periodic decay and revival of entanglement. In addition, the amplitude of each successive rebirth of entanglement decays asymptotically. To conclude, we investigated the entanglement dynamics of single photon coherence and two photon coherence MNMSs and MEMSs under different environments and found that single photon coherence states are more robust against environmental decoherence. This study is expected to be helpful in providing new insights for the development of techniques for experimental realization of information processing tasks using quantum systems.

\section{Acknowledgments}

NN and AP acknowledges the support from Interdisciplinary Cyber Physical Systems (ICPS) programme of the Department of Science and Technology (DST), India, Grant No.: DST/ICPS/QuST/Theme-1/2019/6 (Q46).

\bibliographystyle{ieeetr}
\bibliography{Xstate}

\end{document}